\documentstyle[aps,prl,multicol,epsf,amssymb]{revtex}
\def\gs{\gtrsim}
\def\ls{\lesssim}
\def\gdot{\dot{\gamma}}
\def\be{\begin{equation}}
\def\en{\end{equation}}                  
\newcommand{\bi}[1]{\mbox{\boldmath$#1$}}
\newcommand{\av}[1]{\langle{#1}\rangle}
\begin{document}
\draft
\bibliographystyle{prsty}
\title{Rheology of a Supercooled Polymer Melt}
\author{Ryoichi Yamamoto 
and Akira Onuki}
\address{Department of Physics, Kyoto University, Kyoto 606-8502, Japan}
\date{\today}
\maketitle

\begin{abstract}
Molecular dynamics simulations are performed 
for a polymer melt composed of short chains 
in quiescent and sheared conditions. 
The stress relaxation function $G(t)$ exhibits a stretched 
exponential form in a relatively early stage and ultimately follows 
the Rouse function in quiescent supercooled state.
Transient stress evolution after application of shear 
obeys the linear growth $\int_0^t dt'G(t')$ for strain less than 0.1 
and then saturates into a non-Newtonian viscosity.
In steady states, strong shear-thinning and elongation of chains 
into ellipsoidal shapes 
are found at extremely small shear.
A glassy component of the stress is much enhanced in these examples. 
\end{abstract}


 \begin{multicols}{2}


 Stress and dielectric 
 relaxations of glassy polymer melts occur from microscopic 
 to macroscopic time scales  in very complicated 
 manners \cite{matsuoka,stroble}. 
 Experiments  have shown 
  that the stress relaxation function 
 $G(t)$  exhibits 
 a glassy stretched exponential decay, 
  a glass-rubber transition, 
 a rubbery plateau,   and a terminal decay, in this 
 order over many decades of time. 
 Such hierarchical relaxation behavior arises from 
 rearrangements  of  jammed 
  atomic configurations 
 and  subsequent evolution 
 of chain conformations  
  described by 
 the Rouse or reptation dynamics \cite{doi,kremer}.  
 The stress-optical relation 
  between birefringence and  stress 
  has also been reported to be violated  as the temperature $T$ 
  is approached  the glass transition 
 temperature $T_g$ \cite{muller,muller1,inoue}, 
  obviously owing  to  enhancement  of a glassy part 
 of the stress.

 Recent  simulations on glassy  polymer 
  melts  have mainly 
 treated self-motions of particles 
 in quiescent  states 
 \cite{kopf,binder,bennemann}. 
 However, not enough theoretical efforts 
 have  been made  
 on the rheological properties of
 glassy polymers. 
 Hence, we  will first 
 study  linear rheology of a model short chain system  
 via very long 
 molecular dynamics simulations. 
 Then, we  will demonstrate 
 that  chains are very easily 
 elongated at  extremely small 
 shear rate $\gdot$ 
 on the order of the 
 inverse Rouse time.  
 Marked shear-thinning  
 behavior then takes place for larger shear rates, 
 indicating a decrease of the monomeric friction 
 among different chains.   
 On the other hand, in 
 supercooled  simple fluid mixtures 
  \cite{yo2},   the shear-dependent  
  structural rearrangement   time 
  $\tau_b(\gdot)$ depends 
  on shear as 
  $\tau_b(\gdot)^{-1}= 
 \tau_b(0)^{-1}+ A_b\gdot$,  
 where  $A_b$ is of order 1  
 and $\tau_b(0)$ is on the order of 
 the so-called 
 $\alpha$ relaxation time 
 $\tau_\alpha$ obtained from  the 
 incoherent van Hove time correlation function. 
 The steady state viscosity is expressed 
 in a very simple form,  
 $\eta(\gdot)= A_\eta\tau_b(\gdot)+ \eta_B$, 
 for any $T$ and $\gdot$,    
 where $A_\eta$  and  $\eta_B$ are  constants. 
 We have also found that the 
 cage breakage occurs collectively 
 in the form of clusters characterized by 
 a correlation length $\xi$ \cite{yo1}, 
 where the dynamic scaling 
 $\tau_b(\gdot) \sim \xi^2$  holds 
 in three dimensions.

 In our model  
  all the  bead particles interact 
 with a Lennard-Jones 
 potential of the form \cite{kremer,kopf,bennemann},  
 $U_{LJ}(r)= 
 4\epsilon [(\sigma/r)^{12}-(\sigma/r)^{6}] + 
 \epsilon$. It is cut off 
  at the minimum distance $2^{1/6}\sigma$, 
  so we use its repulsive part only to 
 prevent spatial overlap of particles.  
 Consecutive beads on each chain 
 are connected by  
 an anharmonic spring of the form, 
 $U_{F}(r)= -\frac{1}{2}k_c R_0^2 
 \ln[1-(r/R_0)^2]$ with 
 $k_c=30\epsilon/\sigma^2$ and $R_0=1.5\sigma$, 
 so the bond length cannot exceed $R_0$.  In a cubic box with length 
 $L=10\sigma$ under the periodic boundary condition, 
 we put $M=100$ chains composed of  
  $N=10$ beads.   The number density is 
 fixed at a very high value of 
 $n= NM/V =1/\sigma^3$, which results in   severely 
 jammed configurations at low $T$. 
 We will measure space and time in units of 
 $\sigma$ and  $\tau_0=({m\sigma^{2}/\epsilon})^{1/2}$ with $m$ being 
 the mass of a bead. 
 The temperature $T$ will be measured  in units of $\epsilon/k_B$.   
 Simulations were  
  performed in normal ($T=1.0$) and 
 supercooled ($T=0.4$ and $0.2$)  states 
 with and without shear.  
 The bond lengths 
  $b_{j}= |{\bi R}_{j}-{\bi R}_{j+1}|$ 
 between consecutive beads on each  chain 
 exhibit only small deviations  on the  order of 
 a few $\%$ from  $b_0 \cong 0.96$,  which 
 gives the minimum  of 
 $U_{LJ}(r)+U_F(r)$,  for any $T$ and $\gdot$ in our study.  
 Thus the bond lengths are 
 nearly fixed in our model system.

 We took  data  after long equilibration periods
 ($10^6$ at $T=0.2$) to suppress 
  aging (slow equilibration) 
 effects  in various quantities such as the pressure 
 or the density time correlation functions.
 At zero shear we imposed 
 the micro-canonical condition with the time step 
 $\Delta t = 0.005$.
 In order to obtain accurate 
 linear viscoelastic behavior,  
 very long simulations  of order $10^2 \tau_R$  
 were   performed, where $\tau_R$ 
 is  the primary Rouse relaxation time.  
  In the literature \cite{kremer,kopf,bennemann}, simulation 
 times have been typically  up to $\tau_R$  
 in supercooled states. 
 In the presence of shear 
 we set $\Delta t=0.0025$ and kept  
 the temperature  
 at a constant  using the Gaussian constraint 
 thermostat to eliminate viscous heating.
 After a long equilibration 
 time in a quiescent state  for $t < 0$, 
 all the particles acquired 
   the average flow velocity $\gdot y$ 
 in the $x$ direction  at $t = 0$ 
 and   then the Lee-Edwards boundary condition 
 \cite{Allen,Evans}  maintained   the simple shear flow.  
 Steady sheared states were  realized 
 after transient viscoelastic behavior.

 As has been confirmed in the literature 
 \cite{kopf,binder,bennemann},  
 the dynamics  in  quiescent supercooled states 
 is reasonably well  described by the Rouse model, 
 where   the relaxation time of 
 the $p$ th  mode 
 of a chain is expressed as 
 $\tau_p= \zeta b^2/[12 k_BT \sin^2(p\pi/2N)]$ 
 ($p=1, \cdots, N-1$) \cite{verdier}.  Here, the statistical segment 
  length $b$  is   related to 
 the variance of the end-to-end vector of a chain 
  ${\bi P}= {\bi R}_N - {\bi R}_1$ by 
   $b= [\av{{\bi P}^2}/(N-1)]^{1/2}$, which is 
 $1.17$, $1.18,$ $1.19$ 
 for $T= 1.0$, $0.4$, 
 $0.2$, respectively.  
 We  determined the monomer friction constant $\zeta$ 
 from the relaxation  
 $\av{{\bi P}(t)\cdot{\bi P}(0)}
 = 2N^{-1} \sum_{\ell=0,1,\cdots} \cot^2(\pi (2\ell+1)/2N) 
 \exp (- t/\tau_{2\ell+1})$  of the end-to-end vector.     
 The Rouse relaxation time $\tau_R (=\tau_1 
 \cong \zeta b^2N^2/3\pi^2k_BT)$  
  then increases  drastically with lowering $T$ as 
  $\tau_R=250$, $1800$, and $6\times 10^4$ for $T= 1.0$, $0.4$, 
 $0.2$, respectively. 
 We  also calculated the 
 $\alpha$ relaxation time $\tau_\alpha$ 
 from $F_q(\tau_\alpha)=e^{-1}$ at $q=2\pi$  \cite{kopf},  
 where  $F_q(t)= \sum_{j=1}^N 
 \av{\exp[i{\bi q}\cdot({\bi R}_j(t)-{\bi R}_j(0)]}/N$ 
 is the van Hove self-correlation function 
 for the displacement of a tagged particle.  
 Then we obtained the result 
 $\tau_\alpha \cong 
 0.017 \zeta b^2/ k_BT$ at any $T$, 
 so  we have 
 $\tau_R \cong  2 N^2\tau_\alpha$ in our system.

 Now let us discuss  
 the linear viscoelastic behavior in  
 supercooled states. 
 In Fig.1 we show the stress relaxation function, 
 \be
 G(t) = \langle \sigma_{xy}^T(t)\sigma_{xy}^T(0)\rangle/V k_BT,
 \en    
 where $\sigma_{xy}^T$ 
 is the space integral of the $xy$ component of the total stress 
 tensor over the volume $V=L^3$. 
 At the lowest temperature $T=0.2$, $G(t)$ 
 exhibits salient  features of glassy polymer melts 
 \cite{matsuoka,stroble}.   
 Its initial value is   
 of order 100 (in units of $\epsilon/\sigma^3$) 
 and is very large, and  it 
 relaxes to a  value $G_0$ about 5 for $t \gs 1$. 
 We then have a slow decay of the form,   
 \be
 G(t) \cong  G_0 \exp [-(t/\tau_s)^{\beta}] ,
 \en  
 where $\tau_s=90 \sim \tau_\alpha$ 
 and $\beta=0.5$. 
 The agreement to Eq.(2)  is excellent for
 $1 \ls t \ls 10\tau_s$.  
  This  glassy behavior  arises from monomeric 
 structural relaxation.  
 For   $t\gs 50\tau_s$ 
 it approaches 
 the Rouse stress relaxation function,  
 \be
 G_R(t)=
 n k_B T  N^{-1}\sum_{p=1}^{N-1}
 \exp(-{2t}/{\tau_p}),
 \en 
  which is 
 equal to $T(N-1)/N$ 
 for  $t \ls \tau_\alpha$ and decays as 
 $TN^{-1} \exp(-2t/\tau_R)$ for $t \gg \tau_R$ in the dimensionless 
 units.  Obviously, this  final stage 
  behavior arises from relaxation of  large scale 
 chain conformations. 
 The crossover between these  two regions 
 occurs in a  narrow time time range in our case.  
 Experimentally, however,  the 
   intermediate region, 
   which connects the  glassy and 
   polymeric  (Rouse or reptation)  
   relaxations,  extends over a much wider 
 time  range (typically 4 decades \cite{matsuoka}), 
  and $G(t)$ there 
  has been fitted to an  algebraic  form,  
 $G(t) \cong  e^{-1}G_0 (t/\tau_s)^{-a}$  
 with $a \sim 0.5$ \cite{matsuoka,stroble}.  
 In addition, with increasing 
 the molecular weight,  
 a rubbery plateau 
  has been observed to develop 
 after the crossover 
 before
 \begin{figure}[t]
 \epsfxsize=2.8in
 \centerline{\epsfbox{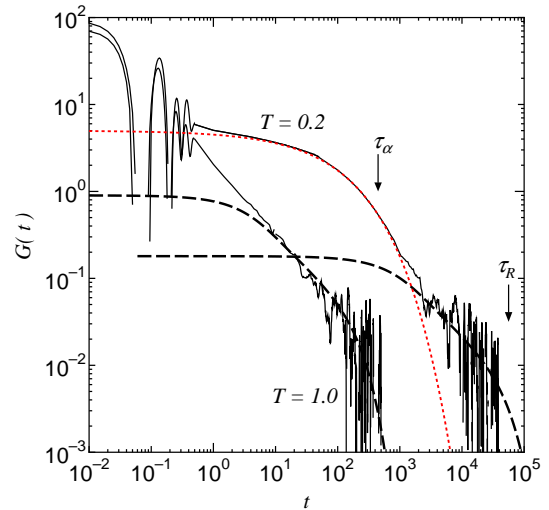}}
 \caption{\protect\narrowtext 
 The  stress relaxation function $G(t)$ 
 (thin-solid lines) at $T=0.2$ in a supercooled state  and 
 $T=1$ in a normal liquid state. 
 It may be  fitted to the stretched exponential form 
 (dotted line) at relatively short times  
 and  tends to the Rouse relaxation function 
  $G_R(t)$ (bold-dashed lines) at long times.
 }
 \label{fig1}
 \end{figure}
 \noindent
 the terminal  decay,  whereas  it is not apparently 
 seen   in our short chain system.  
 In our case, the (zero-frequency) 
 linear viscosity $\eta$ 
  consists of 
 a monomeric part $\Delta\eta$ 
 of order $10\tau_s$ 
 from the integration in the time region 
 $t\ls 10 \tau_s$ 
 and  the Rouse viscosity 
 $\eta_R= \int_0^\infty dtG_R(t)
 \cong 0.808 TN^{-1}\tau_R$ from $t  \gs \tau_R$. 
 The ratio $\Delta\eta/\eta_R$ is 
 thus of order $1/(TN)$ ($\sim 1$ at $T=0.2$), 
 whereas  we should have 
 $\Delta\eta \ll \eta_R$ for much larger $N$. 

 In the Rouse model,  
 the space integral of the polymer (entropic) 
  stress $\sigma^R_{\alpha\beta}$ 
  is the sum of $k_BTb_{j\alpha}b_{j\beta}/b^2$ 
  over all the bonds in the system, where $b_{j\alpha}$ are 
  the Cartesian 
components of  
   the bond  vectors  $
  {\bi b}_j= {\bi R}_{j+1}-{\bi R}_{j}$. 
 We have confirmed that 
 the relaxation function 
 $G_c(t)= \av{\sigma^R_{xy}(t)\sigma^R_{xy}(0)}/Vk_BT$ 
  nearly coincides  with 
  $G_R(t)\cong G(t)$ for $t\gs 0.1\tau_R$, 
   whereas it is about a half of 
   $G_R(t)$ for $t \ls \tau_s$. 
 We note that the bond vectors have the  nearly fixed length 
 $b_0 \cong 0.96$,   and  a bond orientation tensor  
 $Q_{\alpha\beta}$ may be defined as 
 \be
 Q_{\alpha\beta}= (N-1)^{-1}\sum_{j=1}^{N-1}
 \av{b_{j\alpha}b_{j\beta}}/b_0^2. 
 \en  
 Then we have $Q_{\alpha\beta} \propto 
 \av{\sigma^R_{\alpha\beta}}$. 
 If  the electric polarizability tensor of 
 a bead is uniaxial along  the bond vector, 
 the deviation  of the dielectric tensor  
 $\Delta\varepsilon_{\alpha\beta}$ 
 is proportional to 
  $Q_{\alpha\beta}-\delta_{\alpha\beta}/3$. 
 In flow  birefringence we have,  
 \be
 {\Delta\varepsilon_{xy}}= 
 C\av{\sigma^R_{xy}}/V,  
 \en     
 where $C$ is a constant. 
 In supercooled states 
  $\av{\sigma^R_{xy}}/V$ can be 
   much smaller than the total shear 
  stress $\sigma_{xy}$, for instance, in transient states 
  or in  oscillatory  shear. 
 \begin{figure}[b]
 \epsfxsize=2.8in
 \centerline{\epsfbox{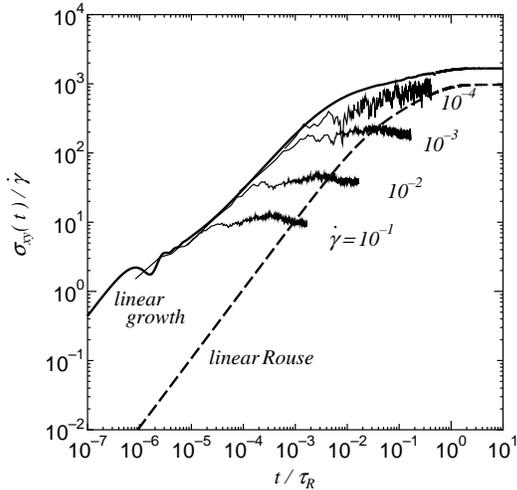}}
 \caption{\protect\narrowtext
 Shear stress  divided by shear rate  
 $\gdot=10^{-1}$, $10^{-2}$, $10^{-3}$, $10^{-4}$ 
 (thin-solid lines)  
 vs $t/\tau_R$ (where $\tau_R=6\times 10^4$) 
 at  $T=0.2$.  
 The curves  follow   the linear growth 
  function (bold-solid line) 
  for $\gdot t \ls 0.1$,  
 but afterwards depart  from it.  The linear 
 growth function in   the Rouse model 
 is also plotted(bold-dashed line).  
 }
 \label{fig2}
 \end{figure}
 \noindent
  Therefore,   the  usual stress-optical law $\Delta\varepsilon_{xy}= 
 C\sigma_{xy}$, which is valid far above $T_g$, 
 breaks  down  close to $T_g$.

 In Fig.2 we show the 
  stress growth function
   $\sigma_{xy}(t)/\gdot$ 
 after application of shear at $t=0$ 
 at the lowest temperature $T=0.2$ for various $\gdot$.  
  The curves are the averages of data of eight 
  independent runs.  
 In the initial stage, in which  $\gdot t \ls 0.1$,  
 we can see the linear viscoelastic growth,  
  $\sigma_{xy}(t)/\gdot=\int^t_0G(t') dt'$,  
  whereas  a nonlinear regime sets in 
  for  $\gdot t \gs 0.1$, resulting in the non-Newtonian 
  viscosity $\eta(\gdot)$.   As a guide, we also plot 
  the linear growth function 
  $\int^t_0G_R(t') dt'$ from the Rouse model, 
  which is much smaller than 
 the true linear growth  for $t \ll \tau_R$.  
   The relevant physical processes 
   are as follows: For $\gdot t\ls 0.1$ 
    the overall chain conformations are 
  affinely deformed, whereas  for $\gdot t\gs 0.1$ 
   the structural rearrangements 
   among beads  belonging to different chains 
  become   appreciably induced by shear 
  (as in the case of supercooled simple fluids \cite{yo2}).  
   Experimentally, a stress overshoot (a rounded  maximum of 
  $\sigma_{xy}(t)$)   has  been observed at $\gdot 
  t =0.05-0.1$ for higher molecular weight 
  melts close to $T_g$ \cite{matsuoka}.

 In Fig.3 we display the steady state viscosity   
  $\eta(\gdot)$ at $T=0.2,~ 0.4,$ and  $1$.  
 It  exhibits marked 
  shear-thinning behavior  for 
 $\gdot\tau_R\gs 1$ and 
  becomes independent of $T$ for 
   very high shear rates. 
 The horizontal arrows indicate the linear 
 viscosity from  the Rouse model $\eta_R$, and the vertical arrows indicate
 the points at which $\gdot=\tau_R^{-1}$. 
 In particular, the curve of  $T=0.2$ may be 
 fitted to  $\eta\propto \gdot^{-\nu}$ with
 $\nu\simeq 0.7$ for 
 $\gdot\tau_R\gs 1$. 
 Similar shear-thinning  has 
 been reported in MD simulations of 
  short chain systems in normal liquid states,  
  but at much higher shear rates 
   \cite{chynoweth,khare}. 
 In the case of supercooled  simple fluid mixtures 
  \cite{yo2}, shear-thinning 
 becomes apparent for 
  $\gdot \gs 0.1 \tau_\alpha^{-1}$. 
 \begin{figure}[t]
 \epsfxsize=2.8in
 \centerline{\epsfbox{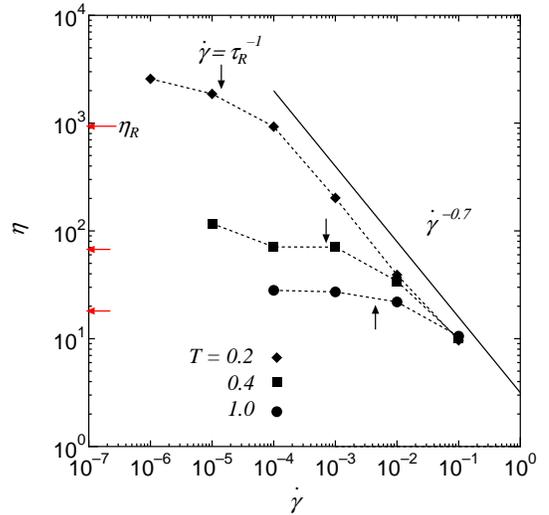}}
 \caption{\protect\narrowtext
 The  steady state viscosity vs $\gdot$ 
 for $T=0.2,~0.4,~1$. A line of slope $-0.7$ is 
 also written as a view guide.
 }
 \label{fig3}
 \end{figure}
 \noindent
 In the present 
 short chain system,  
 significant shear-thinning 
 occurs at much smaller shear 
 of order $\tau_R^{-1}\sim \tau_\alpha^{-1}N^{-2}$. 
 At this early onset,  the 
 overall elongation of 
 chains take place, 
 as will become evident  in Fig.5 below. 
 It is worth noting that 
 the shear stress at 
 $\gdot \sim \tau_R^{-1}$ 
 is of order 
 $nk_BT /N$ (which would be the 
 modulus in the 
 rubbery  plateau for longer chain systems).

 We finally  examine anisotropy 
  in the chain conformations in  shear at $T=0.2$ and $\gdot=10^{-4}$. 
 In Fig.4 (a),  we plot the $xy$ cross section ($z=0$)  
 of the steady state  bead distribution function, 
 \begin{equation}
 g_s({\bi r}) = N^{-1} 
 \sum_{j=1}^N\av{\delta({\bi R}_j-{\bi R}_G-{\bf r})},
 \end{equation}
 where ${\bi R_G(t)}={N}^{-1}\sum^{N}_{n=1}{\bi R}_n(t)$ is the center of 
 mass of a chain. In Fig.4 (b), we plot  
  the  structure factor, 
 \begin{equation}
 S({\bf q})= N^{-2} \sum_{i,j=1}^N 
  \av{\exp(i{\bi q}\cdot({\bi R}_i -{\bi R}_j)} ,
 \end{equation}
 in the $q_x q_y$ plane ($q_z=0$).  
 It  is proportional to 
 the scattering intensity 
 from labeled chains in shear \cite{picot}. 
 We recognize that 
 $g_s({\bi r})$ and $S({\bi q})$ 
 almost saturate into the forms in 
 Fig.4 for $\gdot \gs 10/\tau_R$. 
 The figures indicate that our short 
 chains take  ellipsoidal 
 shapes  on the average once $\gdot\gs 10/\tau_R$. 
 The angle $\theta$ between 
 the  ellipsoids  and the 
 $y$ (shear gradient) direction 
 is also written, 
  which is calculated using Eq.(8) below. 
 Let us define the tensor
 $I_{\alpha\beta} = \sum_{i,j=1}^N \av{
 ({\bi R}_i -{\bi R}_j)_\alpha 
 ({\bi R}_i -{\bi R}_j)_\beta}/N^2$. 
 For small $q$ with $q_z=0$,  we have 
 the expansion,  
 \be
 S({\bi q})
 = 1-   \frac{1}{2}a_1^2 ({\bi q}\cdot{\bi e}_1)^2 - 
 \frac{1}{2}a_2^2 ({\bi q}\cdot{\bi e}_2)^2+\cdots, 
 \en 
 where $\{{\bi e}_1, {\bi e}_2\}$ and  $\{a_1^2, a_2^2\}$ 
 are the eigen unit vectors and values of the 
 tensor $I_{\alpha\beta}$ with $\alpha,\beta=x,y$. 
 The two lengths 
 $a_1$ and $a_2$ are the shorter and longer radii of  the
 \begin{figure}[t]
 \centerline{
 \epsfxsize=1.8in\epsfbox{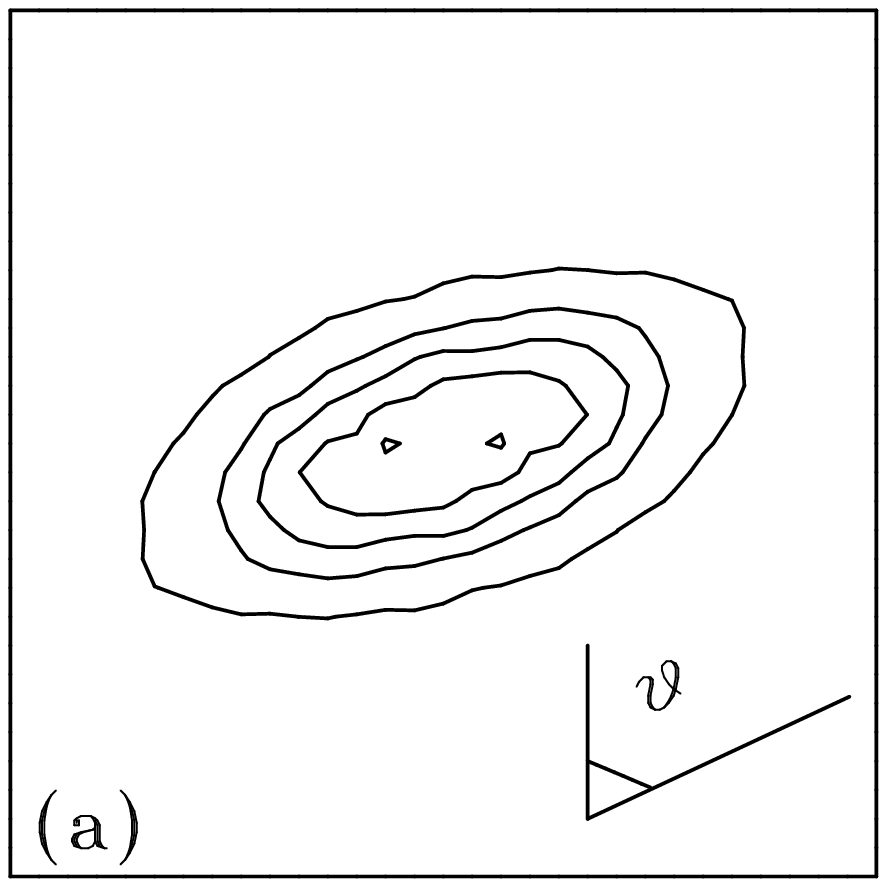}
 \epsfxsize=1.8in\epsfbox{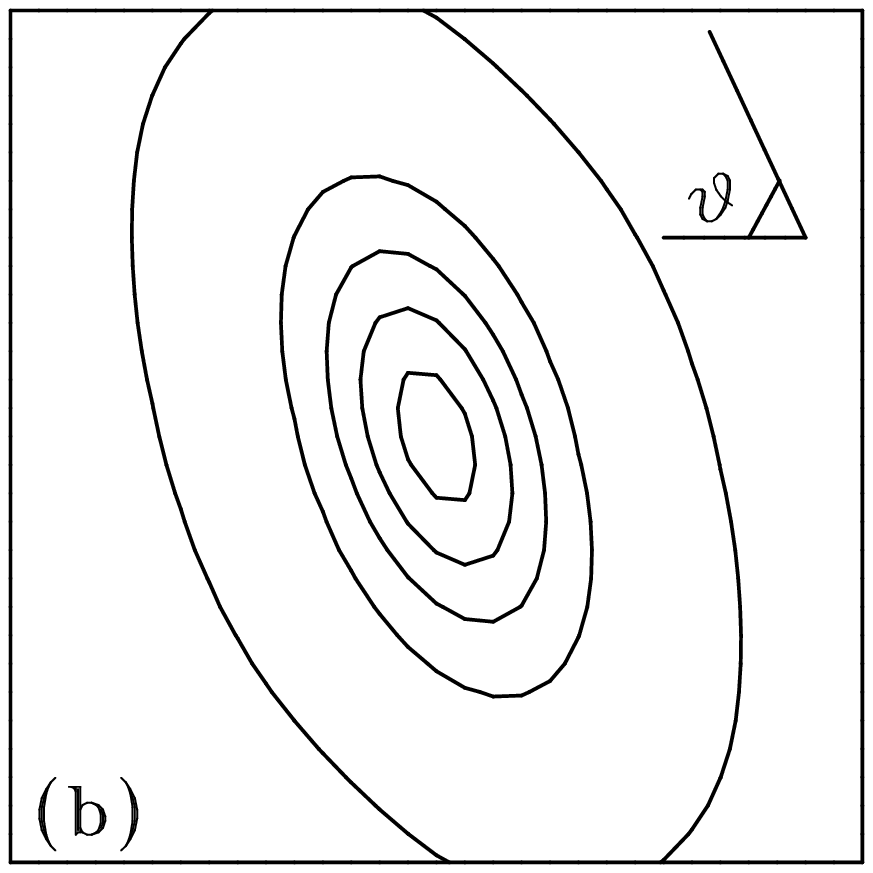}}
 \caption{\protect\narrowtext
 (a) Isointensity 
  curves of  $ g_s({\bf r})$ in Eq.(6) 
 in the $xy$ plane ($-3.75<x,y<3.75,~ z=0$). 
 (b) Those  of 
  the incoherent structure factor 
 $S({\bf q})$  in Eq.(7) 
 in the $q_xq_y$ plane ($-\pi<q_x,q_y<\pi, ~q_z=0$).  
The values on the isolines are $0.01+0.02n$ in (a) 
 and $0.1+0.2n$ in (b) with $n=0,1,\cdots,4$ from outer to inner.
  Here  $T=0.2$,  $\gdot=10^{-4}$, 
and the  flow is  in the horizontal ($x$) direction.
The $\theta$ is the angle between  the average 
chain shapes and  the $y$ axis. 
}
\label{fig4}
\end{figure}
\vspace{-5mm}
\begin{figure}[t]
\epsfxsize=2.8in
\centerline{\epsfbox{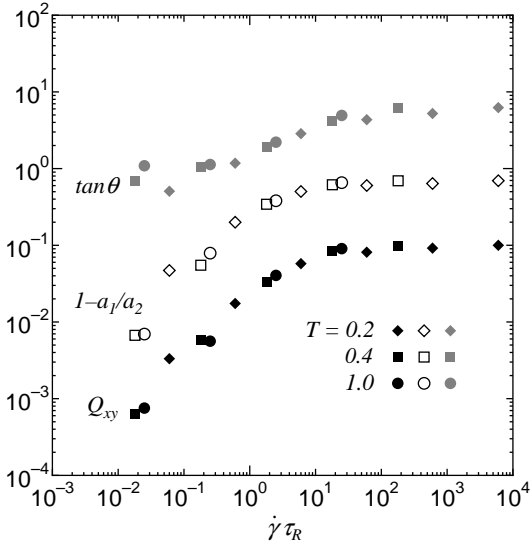}}
\caption{\protect\narrowtext 
$\tan\theta$, $1-a_1/a_2$, and $Q_{xy}$ vs 
$\gdot \tau_R$.
}
\label{fig5}
\end{figure}
\noindent
 ellipses. 
 In Fig.5 we write  
 $\tan\theta= -{ e}_{1y}/e_{1x}$, the degree of elongation   
  $1-a_1/a_2$, 
 and the $xy$ component $Q_{xy}$ 
  of the tensor  in Eq.(4). 
 For $\gdot\tau_R\gs 10$  we have 
 $\theta\cong 80^\circ$, $a_1/a_2 \cong 0.3$, 
 $Q_{xy}\cong 0.1$. 
 These quantities represent the average chain forms 
 and  bond orientation and  
 are  insensitive to $T$ 
 if  plotted vs $\gdot\tau_R$. 
We stress that 
the shape changes  of 
chains start to occur at  $\gdot \sim 
 \tau_R^{-1}$ while  the 
monomeric structural relaxation 
is only slightly affected by shear. In fact, 
 we found  that the 
 cage  breakage time $\tau_b(\gdot)$ 
 among neighboring beads belonging 
 to different chains \cite{yo2} 
 does  not change appreciably 
 for $\gdot \sim \tau_R^{-1}$ at $T=0.2$. 
 This tendency should be more evident  
 for longer chain systems.

 (i) In our simulations, 
 the stress relaxation function 
 obeys a stretched exponential decay and 
 the Rouse relaxation, although 
 the chain length is too short and 
 the temperature is too high  to 
 reproduce the glass-rubber transition 
 region  and the rubbery plateau. 
 (ii) We have also found 
  very early onset of the nonlinear 
 regime of  shear. 
 Strong shear-thinning and 
 anisotropic scattering 
 can  be predicted 
 for  $\gdot \gs \tau_R^{-1}$ 
 or for $\sigma_{xy} \gs nk_BT/N$ 
 in  the Rouse  case $N<N_e$. In  the entangled case     $N>N_e$   
 the threshold shear rate and  stress 
 needed for the onset of nonlinearity  
 should be of order   $\tau_{rep}^{-1}$ 
 and $nk_BT /N_e$, respectively, where 
 $\tau_{rep}$ is the reptation time. 
 Scattering experiments from 
 very weakly sheared 
 melts near $T_g$ 
 are  promising.   
 (iii) Although  not presented here, 
  heterogeneities are much enhanced at low $T$ 
 in the cage breakage events 
 among beads belonging to 
 different chains \cite{yo1}. 
 They are  characterized by the correlation length 
 $\xi$ dependent on $T$ and $\gdot$. 
 Interesting crossover effects can then 
 be expected at $\xi \sim bN^{1/2}$ for longer  chain 
 systems.

 This work is supported by Grants in Aid for Scientific 
 Research from the Ministry of Education, Science, Sports and Culture
 of Japan.
 Calculations have been performed 
 at  the Human Genome Center, Institute of Medical Science,
 University of Tokyo.

\end{multicols}
\end{document}